# Application of Visual Clustering Properties of Self Organizing Map in Machine-part Cell Formation

Manojit Chattopadhyay[1], Pranab K. Dan[2], Sitanath Majumdar[3]
[1] Department of Computer Application, Pailan College of Management & Technology, Kolkata-700104
[2] Industrial Engineering, School of Engineering & Technology, West Bengal University of Technology, Kolkata 700 064
[3]Department of Business Management, University of Calcutta, Kolkata- 700 027
Email: [1] chattomanojit@yahoo.com,
[2] dan1pk@hotmail.com
[3] sitanath_mazumdar@rediffmail.com

## ABSTRACT

Cellular manufacturing (CM) is an approach that includes both flexibility of job shops and high production rate of flow lines. Although CM provides many benefits in reducing throughput times, setup times, work-in-process inventories but the design of CM is complex and NP complete problem. The cell formation problem based on operation sequence (ordinal data) is rarely reported in the literature. The objective of the present paper is to propose a visual clustering approach for machine-part cell formation using Self Organizing Map (SOM) algorithm an unsupervised neural network to achieve better group technology efficiency measure of cell formation as well as measure of SOM quality. The work also has established the criteria of choosing an optimum SOM map size based on results of quantization error, topography error, and average distortion measure during SOM training which have generated the best clustering and preservation of topology. To evaluate the performance of the proposed algorithm, we tested the several benchmark problems available in the literature. The results show that the proposed approach not only generates the best and accurate solution as any of the results reported, so far, in literature but also, in some instances the results produced are even better than the previously reported results. The effectiveness of the proposed approach is also statistically verified.

## 1. Introduction

In the competitive global market, the survival of even the most well-established world class manufacturing firms depends on the ability to stay flexible and at the same time to reduce operating costs. Cellular Manufacturing System (CMS) is an application of Group Technology (GT) which helps firms to reach this goal. The group technology (GT), proposed by Mitrafanov (1966) and Burbidge (1963, 1979), seeks to identify and group similar parts to take advantage of their similarities in manufacturing and design as well as to achieve economy in production. One application of GT is cellular manufacturing (CM). CM is concerned with processing a similar part in a part family by a dedicated group of machines or processes in a cell (Tompkins et al, 1996). The benefits reported in the literature on the CMS are reduction in throughput times, setup times, work-in-process inventories, and response times to the customer orders, etc. (Wemmerlov and Hyer, 1987; Wemmerlov and Johnson, 1997).

The cell formation problem (CFP) in the production flow analysis approach is a machine-part incidence matrix (MPIM) in the form A= [$a_{ij}$] which is a $m \times p$ matrix. If $a_{ij}$ =1 then it means that machine *i* processes the part *j*, otherwise $a_{ij}$=0. Each binary element in the $m \times p$ matrix indicates a relationship between machines and parts where ''1'' or ''0'' represents that the pth part should be processed on the mth machine or otherwise. The matrix also displays all similarities in parts and machines. The objective is to group parts and machines in a cell based on their similarities. If there are no ''1'' outside the diagonal block and no ''0'' inside the diagonal block then it is termed as perfect result.

The binary MPI matrix is represented by '1' whereas the sequence of operation of the parts visiting the respective machines are represented with 1, 2, 3 etc and the '0' will remain same with the meaning that there is no visit. In spite of a large number of contributions made towards various cell formation solution methodologies applied on binary (considering only ‘1’s in the input matrix) MPIM it appears only few of them reported findings with grouping objectives using operation sequence till date. This operation sequence data provide important information related to the flow patterns of different jobs in a manufacturing system (Mahdavi et al., 2008). Although Suresh et al., (1999) reported that research on sequence based clustering is in an incipient phase. Several authors, for example Dixit and Mishra (2010), Bu *et al*, (2009), Pandian and Mahapatra (2009); Chan *et al*, (2008), Kumar and Jain (2008), Jayaswal and Adil (2004), Sarker and Xu (1998), Tam (1990), Vakharia and Wemmerlov (1990), Choobineh (1988). Sarker and Xu (1998) reported that there are primarily four types of methodologies that have been applied in solving operation sequence based cell formation, viz., mathematical programming, network analysis, materials flow analysis method, and heuristics. Mahdavi and Mahadevan (2008) reported that sequence based cell formation is the least explored area although sequence data provides valuable information about the flow patterns of various jobs in a manufacturing system.

Wang (1999) proposed the number of feasible solutions ($n_{soln}$) of the CFP as

$$n_{soln} = \left( \left( \sum_{i=1}^{p} (-1)^{p-i} \; i^{n} / \left[ i!(p-i)! \right] \right) * \left( \sum_{i=1}^{p} (-1)^{p-i} \, i^{m} / \left[ i!(p-i)! \right] \right) \right) \qquad (1)$$

The above equation (1) shows the number of feasible solutions ($n_{soln}$) of the CFP based on the number of probable ways n parts and m machines can be grouped in p clusters (assuming there are p clusters for the CFP). Currently the researchers are being attracted more to the unsupervised artificial intelligence and meta-heuristic approaches as evidenced from literature due to consideration of this number as computationally infeasible (Srinivasulu and Jain, 2006; Ravi et al., 2008).

Extensive reviews of the works involving the finding of optimum machine-part cluster using mathematical programming, heuristic and metaheuristic and artificial intelligence as strategies have been reported for the solution of the cell formation problem (see Papaioannou and Wilson, 2010; Singh and Sharma, 2006; Rezaeian et al.,2011 and references therein). Different optimization solution of a number of design approaches for various CFP has been studied to develop a taxonomic framework (Yin and Yasuda, 2006).

In view of the literatures surveyed above, the present work endeavors to apply an unsupervised competitive learning algorithm, such as the self-organizing map (SOM) (Kohonen, 1990, 1998, 2001) to visualize machine-part clustering based on operation sequence data. The visual clustering of SOM performs both the vector quantization and the vector projection for which it becomes an effective tool (Flexer, 2001).

A number of intelligent applications developed by SOM are reported so far (e.g. Abonyi et al., 2003; Heikkinen et al., 2009; Jämsä-Jounela et al.,2003; Kohonen et al., 1996; Liukkonen et al., 2009; Liukkonen et al., 2011). The use of the SOM for different purposes has produced a diversified range of applications in different practical areas, including exploratory data analysis (Hébrail et al., 2010), pattern recognition (Khosravi and Safabakhsh, 2008; Mari et al., 2010) speech analysis (Moschou et al., 2007), industrial and medical diagnostics (Yu, 2011;Pandit et al., 2011; Zhang et al., 2010), robotics and instrumentation (Johnsson, and Balkenius, 2010 and 2011; Yamada, 2004), and even control (Kohonen, 2001, Oja et al., 2002). Moreover, Chang and Liao(2006), Corona *et al*., (2010), Fuertes et al., 2010; Kasslin et al., 1992 and Alhoniemi et al., 1999 have introduced SOM-based applications to process monitoring and modeling in many industrial cases. The cluster visualization has been shown to successfully implement in multidimensional data (Fernandez and Balzarini, 2007).

Furthermore, recent studies like Chu (1993); Venugopal and Narendran (1994); Kiang et al (1995); Kulkarni and Kiang (1995); Inho and Jongtae ( 1997); Kuo et al., 2001; Kiang, 2001; Ampazis and Minis (2004); Venkumara and Haq (2006); Chattopadhyay et al., 2011 have reported the power of SOM in the modeling of the machine-part cell formation in cellular manufacturing. But the solution approach using SOM to the cell formation problem based on operation sequence are rare in the literature. The limitation of the majority of the available works is that the machine-part cell formation is not generated simultaneously. The novelty of the present work is to analyze the measures of SOM quality for cluster visualization during machine-part cell formation. In this paper we explore the efficiency of modeling of the machine-part cell formation using the SOM with the analysis of the visual clustering properties and compare the result with the methodology presented in (Pandian and Mahapatra, 2008 & 2009). A comparison with other method (ART1) has been made to examine its effectiveness of the proposed method and an analysis is presented using an appropriate statistical test. The objective of the present work is stated as follows:

1. It has explored the effect of SOM map unit size on the clustering quality of machine part cell formation.
2. It has analyzed the overall impact on the performance of the machine part clustering due to topographic error (TE), quantization error (QE) and average distance measure of SOM.
3. Statistical verification of the group technology efficiency of the output result obtained using the proposed method.

4. The impact of SOM map unit size, as a metric, has been obtained from corresponding machine-part matrix size on CPU time.
5. The impact of computation time with the different SOM map unit size has also been studied for different types of cell formation problems on the performance measure of machine-part cell formation.

Thus in our present work, we endeavor to construct SOM cluster output that are both good with respect to SOM quality measures and cell formation performance measure.

The organization of the rest of the paper is as follows: in the immediate next section 1.1 an example of CFP based on operation sequence has been elaborated. Issues related to the self organizing map algorithm, visual clustering and SOM quality measures have been discussed in the section 2. It has also discussed about the SOM map size. The section 3 has provided a discussion on performance measure on cell formation. A numerical example has been discussed in section 4. A discussion on computational results appears in section 5. Finally section 6 concludes the paper.

### 1.1 An Example of Cell Formation Problem based on Operation Sequence

An example of machine-part matrix based on operation sequence (ordinal values) is shown in the Table 1. This represents the input matrix derived from the production flow consisting of six machines and eight parts and 24 activities. There are three operations for the parts. Table 2 shows the output block diagonal form with two cells. The two cells are completely independent where each part family (p3, p6, p1, p5, and p2, p8, p4, p7) will be processed only within a machine group (m2, m6, m4 and m1, m5, m3). The order of operation sequence varies for each part in different machines. An optimal result for a machine/part matrix by a cell formation clustering method is desired to satisfy the following two conditions: (a) to minimize the number of 0s inside the diagonal blocks (i.e., voids); (b) to minimize the number of ordinal sequence values outside the diagonal blocks (i.e., exceptional elements). In the cell formation problem identification of optimum number of cluster is done based on similarities of machine-part.

## 2. Overview of Self Organizing Map

The universal application of a self-organizing map (SOM) [Kohonen, 2001] is for mapping $n$-dimensional input vectors to two dimensional *neurons* or *maps*. This map describes variations in the information of the input data, and the topology of the original data is preserved on the SOM through the connection of the input vectors which shares common features of the similar or neighboring neurons.

### 2.1 SOM Algorithm

Typically, the SOM algorithm performs a number of successive iterations until the reference vectors $\boldsymbol{m}_i$ associated to the neuron of a two-dimensional map represent, as much as possible, the input patterns $\boldsymbol{x}_i$ known as Vector Quantization (VQ) that are closer to those neurons. VQ is strongly related to clustering. Clearly, the SOM performs VQ as the sample vectors are mapped to a number of reference vectors. Finally, each sample in the data set is mapped to one of the map neuron which is known as Vector Projection (VP). The VP aims at reducing high dimensional input space to a low dimensional output space and mapping vectors in two dimensional outputs for visualization. The map size can be different based on the type of application. The bigger size map reveals more details of information whereas a smaller map is being chosen to guarantee the generalization capability. The competitive learning algorithm applied to the network (Kohonen, 2001; Kaski, 1997):

An input vector is shown to the network; the Euclidean distances between the considered input vector $\boldsymbol{x_i}$ and all of the reference vectors $\boldsymbol{m}_i$ are calculated.

$$x_i = [x_1, x_2, \ldots, x_n]^T \in \Re^n, i = 1,2,\ldots,N \qquad \ldots(2)$$

$$m_i = [m_1, m_2, \ldots, m_n]^T \in \Re^n, i = 1,2,\ldots,N \qquad \ldots(3)$$

In the training process, the best matching unit (BMU) $m_c$, the unit whose reference vector has the smallest *n*-dimensional Euclidean distance to the input vector, is chosen according to:

$$\|x - m_c\| = \min_i (\|x - m_i\|) \qquad \ldots(4)$$

Simultaneously, the nearest neighbors of the BMU are activated according to a neighborhood function (e.g., Gaussian distribution). Finally the reference vectors of all activated units are updated. The factor hci(t) controls the rate of change of the reference vectors and is called the learning rate.

$$m_i(t+1) = m_i(t) + h_{ci}(t)[x(t) - m_i(t)] \qquad \ldots(5)$$

Where t denotes the index of the iteration step, x(t) is the vector-valued input sample of x in the iteration t. Here, the hci(t) is called the neighborhood function around the winning node c. During training, hci(t) is a decreasing function of the distance between the i-th and c-th model of the map node. For convergence it is necessary that $h_{ci}(t) \rightarrow 0$ when $t \rightarrow \infty$. Therefore, the SOM learning is gradually changing from an early rough learning stage with a large influence neighborhood and fast-shifting prototype vectors to a fine-tuning stage with little neighborhood radius and prototype vectors that adapt gradually to the samples.

**2.2 SOM Cluster Visualisation**

The SOM is an extremely versatile tool for visualizing high dimensional data in low dimension. Dimension reduction of SOM is followed simultaneous vector projection in lower dimension space (Vesanto, 1999). For visualization of SOM map both component planes and the U-Matrix are used which take only the prototype vectors and not the data vector. Followings are the SOM visualization tools:

2.2.1 U-matrix

Similar to all other clustering algorithm the SOM is not an exception of the common problem of deciding boundaries of the clusters (Hautaniemi et al., 2003). In their simple experiment, Ultsch and Siemon(1989), reported that clusters cannot be detected in a reliable approach using the two-dimensional display only. To provide an easy and straightforward detection of the clusters they proposed a unified-distance matrix (U-matrix). The U-matrix calculates distances between neighboring map units, and these distances can be visualized to represent clusters using a grey scale display on the map (Kohonen, 2001).The structure in the data can be seen by the U-matrix which indeed, is the most used tool to visualize patterns by SOM (Skupin & Agarwal, 2008, p. 13).

U-matrix representation of a sample data is shown in Figure 1. The U-Matrix technique (Chattopadhyay et al., 2011) is a single plot that shows cluster borders according to dissimilarities between neighboring units. The distance ranges of U-Matrix visualized on the map lattice are represented by different colors (or grey shades). Dark shades correspond to large distance, i.e. large gaps exist between the code vector values in the input space; light shades correspond to small distance, i.e. map units are strongly clustered together. U-matrices are useful tools for visualizing clusters in input data without having any priori information about the clusters (Tas, 2008).

### 2.2.2 Component planes

Another important tool to visualization is component plane which represent the distribution of each variable on the map grid through the variation of color. Thus it helps to analyze the contribution of each variable to cluster structures of the trained SOM and each input variable (component) calculated during the training process was visualized in each neuron on the trained SOM map in grey scale. The combined use of u-matrix and component planes (Figure 1) are valuable to the researchers as they provide important information about visualizing cluster structure and correlation among variables (Chattopadhyay et al, 2011).

## 2.3 Map Quality Measure

In the literature of SOM there are a number of measures viz., Quantization Error (QE), Topographic Error (TE), SOM Average Distortion Measure (ADM) with two evaluation criteria: measuring the quality of the continuity of mapping and topology preservation or mapping resolution. These measures have been proposed for computing and comparing the quality of a SOM's projection (Kohonen, 2001; Pölzlbauer, 2004). The purpose of all of them is to measure how better the topology is preserved in the projection onto the SOM's lower dimension map grid. Lower QE and TE values specify superior mapping quality.

### 2.3.1 Quantization Error

The quality of learning and fitting of the map can be well understood by the QE. In the present work, first a quantization error computed (Kohonen, 2001) as shown in the equation 6 which is calculated as the average distance between the *m* input patterns □□ and the reference vector □□ associated to their Best Matching Units.

$$QE = \frac{\sum_{i=1}^{m} d(x_i, m_c)}{m} \qquad (6)$$

Where,

$m_c$ is the reference vector associated to the BMU of $x_i : c = \arg\min_j \{d(x_i, m_j)\}$ (7)

$d(x_i, m_c)$ is the Euclidean Distance

m is the number of input patterns

This QE indicates the quality of learning of a SOM and gives an idea how the map fits to data. But a low QE may be associated with an over fitted model (Alhoniemi, *et al.*, 2002).

By increasing the number of the map neurons for any specified data set the QE can be reduced as the data samples are scattered more finely on the map.

2.3.2 Topographic error

Topographic error represents the proportion of all data vectors for which first and second BMUs are not closest to measure the topology preservation (Kivilnoto, 1996).The TE, represents the accuracy of the mapping in the preserving topology (Kohonen, 2001). The topographic error *TE* can be computed in the map as follows (Kiviluoto, 1996):

$$TE = \frac{1}{N}\sum_{k=1}^{N} u(x_k) \qquad (8)$$

where N is the number of input samples, and $u(x_k)$ is 1 if the first and second BMUs of $x_k$ are not next to each other, otherwise $u(x_k)$ is 0.

2.3.3 Average Distortion measure (ADM)

If the data set is discrete and the radius of the neighbourhood kernel is constant then there exists a measure of SOM distortion $E_d$ (in equation 9) which is a local energy function or cost function of the SOM ( Lampinen and Oja, 1992). This function can be used to calculate an error value for the whole map. One of the major advantages is that the error can be decomposed into three parts (Vesanto et al, 2003):

- local data variance –it gives the quantization quality
- neighbourhood variance-assesses the trustworthiness of the map topology and
- neighbourhood bias - links the other two and can be seen as the stress between them.

All of these measures can be used to compare SOMs and (to varying success) to select the SOM that best fits the data. However, none of these measures compares two SOMs with each other or directly shows their differences.

$$E_d = \sum_{i=1}^{n} \sum_{j=1}^{m} h_{b_{ij}} \left\| x_i - m_j \right\|^2 \qquad (9)$$

2.4. Map size

Specification of map size (number of output neurons) in the SOM training process is very much important to detect the deviation of the data. If the map size is too small, outcome is more general patterns (Vesanto et al., 2000) and it may not reveal some significant differences that should be detected. On the other hand, larger map sizes outcome into more detailed patterns where the differences are too small (Wilppu, 1997). Thus, controls run is repeated by changing the map size only during the training of the network, and select the optimum map size where the QE and TE resulted with minimum values.

All experiments, in the present study were performed using the SOM Toolbox (Vesanto et al., 2000) under a Matlab (version 7.6) software (Mathworks Inc., Natick, MA, USA, 2008) platform. The parameters of the SOM and the size of the map were determined by experimental testing.

## 3. Performance Measure

In order to evaluate the performance of goodness of the block diagonal form of output matrix in cell formation problem based on the operation sequence (i.e., ordinal data) using the proposed SOM algorithm, popular measures like group efficiency and group efficacy (Kumar and Chandrasekharan, 1990) cannot be adopted as they are only suitable with binary data but not with the ordinal data. Therefore, a measure known as Group Technology Efficiency (GTE) developed by Harhalakis et al., (1990) can be used to evaluate the performance of cell formation based on sequence of operation. Where the GTE is defined as the ratio of the difference between the maximum number of inter-cell travels possible and the numbers of inter-cell travels actually required by the system to the maximum number of inter cell travels possible as given in the equation 10. The maximum numbers of inter-cell travels possible in the system is

$$I_p = \sum_{j=1}^{N} (n_o - 1) \quad (10)$$

The number of inter-cell travels required by the system is

$$I_r = \sum_{j=1}^{N} \sum_{w=1}^{n_o - 1} t_{njw} \quad (11)$$

GTE is calculated as

$$GTE = \frac{I_p - I_r}{I_p} \quad (12)$$

Where, $I_p$ =maximum number of intercell travel possible in the system

$I_r$ =number of intercell travel actually required by the system

n= number of operations (w=1,2,3,…,n)

$$t_{njw}\begin{cases} = 0 \;\; if \;\; the\, operations\; w, w+1\, are\; performed\; in\, the\; same\; cell \\ = 1 \;\; otherwise \end{cases}$$

The GTE is powerful performance measure as it considered the sequence in which the operations are performed rather than the class in which they are performed and also provides a penalty of 1 for each inter-class travel. The more the value of GTE the better the goodness of cell formation based on operation sequence.

## 4. Numerical example

A number of instance problems, all solved with the proposed visual clustering of SOM, are presented in this section. Example 1 is explained in detail for its input data and computational results. While, other example problems are comparable to example#1, only summarized outcomes are presented to demonstrate the machine-part clustering issues addressed with the proposed model.

A 6x8 dimension machine-part incidence matrix based on operation sequence (ordinal data) has been artificially generated (example#1). The example#1 becomes input (Table 1) for unsupervised training using SOM algorithm. The visual clustering output of SOM for example#1 has been illustrated using the umatrix, component plane. At the end of the learning procedure, areas of neurons with small distances between their weight vectors have been developed on the map. The network arranges itself around the present clusters while recognizing the distribution and the topology of the data in the input space iteratively (Kohonen, 1984; Haykin, 1999; Ghaseminezhad and Karami, 2011). In order to find the best network topology, we applied quantitative criteria by slightly increasing the number of neurons. Several SOM networks with various numbers of nodes and topologies were tested to find the most fitting SOM for the given problem. In solving this example, we consider 6 different types of machines, 8 part types, and three operations performed on these 8 part. The input data of this example are given in Tables 1.

3.2 Finding the best model

Three maps were trained on the data using three different SOM map size (7X7, 10X10, 12X12). Then the three maps were compared with all three visualisations to find out the projection of data vectors, and the similarity and dissimilarity of the map units. Around three models were assessed for quality of vector quantization and map projection based on minimum of qe, te and adm with an acceptable error. The optimum map size of 7X7 is determined after selecting the minimum values of qe (0), te(0) and adm as 0, 0 and 0.393 respectively. The completion time required for the training of the SOM model for this example#1 for the optimum map size of 7X7 comes out to be 2 seconds. The optimum map size based on the criteria of the minimum values of qe, te and adm are presented and summarized in table 4 after assessing 46 SOM models for all the 15 problems.

The unified distance matrix (U-matrix) was used to visualize the SOM accumulations on the resulting two-dimensional feature map (Figure 1). Here, a large distance between the neurons is indicated by a darker shade of the connection between the neurons, which can be interpreted as a class border. The lighter areas indicate a small distance between the neurons (i.e. their weight vectors), indicating closer proximity to the weight vectors in the input space that is spanned by CFP vectors based on sequence data. Hence, light shaded areas can be interpreted as clusters. Figure 1 shows the visualized U-matrix of the map network.

From the U-matrix it is easy to see that the top six rows of the SOM form a very clear cluster (red marked cluster-I). By looking at the labels, it is immediately seen that this corresponds to the parts p2 and p7. After a closer look on the problem data it was found p4 and p8 are absolutely similar to p2 and p7 respectively for which only p2 and p7 have extracted on map. The two other parts p1 and p6 extracted on the other cluster region (red marked as- II) form the other cluster. After studying the problem data then we found p3 and p5 are absolutely correlated with p1 for which only the p1 and p6 are extracted on this cluster region of the u-matrix of SOM map. The u-matrix thus represents two different parts of the cluster. From the component planes it can be seen that the machine m1, m3 and m5 are very much related to each other and projecting on the region of cluster-I of u-matrix of SOM map. On the other hand the machines m2, m4, m6 have extracted on their respective component plane of SOM map in a visually similar way. Moreover, they correspond to the cluster region-II of the u-matrix based on their visual projection as extracted in the respective individual component plane. Component planes of SOM map are very much helpful in visual clustering as it provides detail information on the data structure of the machine variable. Thus cluster-I corresponds to the machine group by the machines m1, m3, m5 and the part family by the parts p2, p4, p7, p8 as machine-part cell-1. While the machine-part cell-2 is formed from the cluster-II which represents the machine group by the machines m2, m4, m6 and part family by the parts p2, p4, p7, p8.

The GTE based on equation 12 is found to be 81.25% . The GTE has been calculated from the block diagonal form of Table 2.

## 5. Computational Results

To evaluate the performance of the proposed algorithm, we tested the fourteen benchmark problems available in the literature. The SOM algorithms were executed using MATLAB 7 to run on a Windows XP with Intel® dual-core technology and 3.0 GHz processor. The sources and input parameters for these problems are shown in Table 3. To cover different sizes, problems with small size (e.g. 5X4), medium size (e.g.20X20) and large size (e.g. 55X20) have been selected. For simplicity, minimum utilization is the same for all cells in each problem as shown in Table 4. A summary of all SOM parameters used for solving the benchmark problems are given in Table 4.

After trial and error experiment for each CFP problem the best optimum map size has been decided based on the criteria of minimum values of qe, te, and adm respectively. Thus in the Table 7, the gray scaled marked rows are the optimal SOM map size where the clustering and topology preservation are the best one. As shown in Table 7, the QE declines as the SOM map size becomes larger. Consequently, the QE cannot be used to contrast maps of different sizes. In three problems (problem#3, 9, 14) in Table 7 , the TE values are more than zero but at the same time in problem#3 and 14 these values converged with minimum values of 0.05, 0 .02 and in problem#9 the criteria of choosing the map size is zero value of qe when te is 0.53. Based on these error values the map size was selected as 49 (7X/7) units for the example#1 dataset. Similarly the SOM map sizes of other data sets have been selected as given in Table 7. Although according to authors of the present paper te is not an impressive criteria of choosing the SOM map size. As shown in Table 7, the value of SOM distortion measure (adm) increases for larger maps. The adm is not an appropriate measure to decide the correct size of the map, but for comparing maps of the same size. The Figure 4 shows that both errors (qe and te) are highest for the problem# 15 with 55x20 dimension for SOM map size 34X34. The Figure 5 shows that, irrespective of the te which is more than zero there is still increase in GTE even for the large size SOM map when the qe is either zero or slightly more than zero (in 11 problems qe or te is zero and in 9 problems both errors are zero).

Table 7 shows the clustering quality of the proposed solution using quantization error (QE), topographic error(TE). The Figure 3 shows the comparison of percentage increase in GTE using proposed solution with the computation time taken to execute the proposed solution for all the different sizes of the 14 problem data sets. Even after higher increase in computation time there is still an increase in performance (GTE) for the big size problem data sets. The SOM clustering measures of performance using TE and QE have been shown in figure 4. Interestingly, both the error measures are diminished with the increase in size of the map or number of SOM nodes alongwith marginal increase of both the measures in small size problems.

The group technology efficiency results of the proposed SOM model in Table 3 are compared with the ART1 method (Sudhakar and Pandian, 2009), which according to its authors generates the best results in the literature for the CFP.

As evident in Table 3, in all the benchmark problems, the group technology efficiency of the solution obtained by the proposed method is either better than that of ART1 method or it is equal to the best one. The Figure 2 shows the significant increase in GTE with respect to GTE even in increase with the size dimension of the problem of both compared approach for each of the benchmarked problem. The SOM algorithm found better solutions in 10 instances (71.42%) than ART1 in instances problem#5 to 14 whereas in all other 4 cases (28.58%) it is as good as the literature one. In other words, the proposed method outperforms the ART1 method and the best solution is therefore being reported in this paper. In four problems (28.57%), namely problem#1to 14, the solution obtained by the proposed method is as good as the best solution available in the literature. In four problems, namely 5, 7, 11, and 13, we have obtained an increase in group technology efficiency more than equal to 10%.

In order to further elaborate the effectiveness of the new approach, a statistical analysis has also been carried out. The details are given in Table 5 and Figure 7a and b. A Paired t-test and confidence interval (CI) of 95% for the mean difference on best result of GTE as obtained from literature (Pandian and Mahapatra, 2009) as well as on the proposed SOM approach are performed. The confidence interval for the mean difference between the two techniques does not include 'zero', indicating a difference between them. The small p-value ($p = 0.035$) further suggests that the data is inconsistent with H0: $d = 0$, that is the two techniques do not perform equally. Specifically, proposed SOM approach (mean = 84.83) performed better than the literature approach (mean = 79.03) in terms of finding out GTE with the extent of the fourteen problems from the literature (Tables 3).

Both GTE from literature and proposed approach results are normally distributed (Anderson-Darling normality test at significance level $\alpha=0.05$; $P=0.927$ for GTE from literature and 0.292 for GTE from proposed SOM approach).

The graphical output in figure 7a and 7b are plots of normal probabilities versus the data. The data depart from the fitted line most evidently in the extremes, or distribution tails. Since in any t-test the assumption of normality is of only moderate importance, therefore, even though the data looks like departing from the fitted line in the lower extreme, still the Anderson Darling (AD) tests' p-value indicates that it is safe applying the Paired t-test. The results obtained in Table 6 under test of equal variances (Levene's test) also support the above statement before a comparison of means.

## 6. Concluding Remarks

The present research introduced an unsupervised self organizing map algorithm for visual clustering to solve the machine-part cell formation problem based on sequence data (ordinal). The self organizing map algorithm combines a clustering through dimension reduction and an efficient visualization of clustering to achieve better group technology efficiency. Using the same representation of a solution, we also have proposed self organizing map quality measures of quantization error, topography error and average distortion measure. We also have analysed computation time during the training of the proposed algorithm as another function of quality of solution.

We compared results of the ART1 algorithm available in the literature with the results of proposed self organizing map algorithms. Our experiments demonstrated the importance of the visual clustering of the self organizing map that generates good quality solutions, as well as the need for a self organizing map quality measures.

We have seen (Figure 3) that even after increase in computation time for the larger size problem (and larger SOM map size) there is still increase in performance of cell formation (Group Technology Efficiency). It also shows that moderate size problems (with average SOM map size) resulted with higher percentage increase in GTE in comparison with low computation time.

To recompense the extra computing efforts required by this proposed approach, we showed that it is adequate to optimize the number of self organizing map size. This is done after evaluating the SOM quality measures of QE, TE, and ADM as minimum for a set of initial map size to decide an optimum SOM map size as described in Table 7 which has been applied to cluster the machine-part cell formation for the 14 problems as shown in the figure 4 and 5. It has been revealed that (Table 7) the TE, QE are almost zero for the best optimum SOM map size when the ADM are smaller for all the benchmark problems.

The quality of the trained SOM has been evaluated by the quantization error and topographic error calculated for various map sizes. Based on this assessment the optimum map size has been determined for all the problem data sets under consideration. It has been discerned that the SOM was smoothly trained in topology shown in Table 7. It is further observed pictorially (see figures 4 and 5) that an optimal map exists for a given input data set. This study further observed that although there are no rules to define the optimal map size, the size of the SOM map has a strong influence on the quality of the classification. To be more specific, it may be stated from figure 6 that an increase in the map size brings more resolution into the mapping. This indicates that the increase in SOM map size is directly related to the low value of TE. However, in this case there is increase in computation time. Therefore, this study concludes that the selection of the initial map size plays an important role in this regard. Comparing the outcomes of the study reported in this paper with the existing ones by means of paired t-test it has been found that the proposed SOM approach performed better than the approach available in the literature.

## References

1. Abonyi J., Nemth S., Vincze C., and Arva P., Process analysis and product quality estimation by self-organizing maps with an application to polyethylene production, Computers in Industry, vol. 52, no. 3, pp. 221–234, 2003.
2. Alhoniemi E., Hollmén J., Simula O., and Vesanto J., “Process monitoring and modeling using the self-organizing map,” Integrated Computer-Aided Engineering, vol. 6, no. 1, pp. 3–14, 1999.
3. Alhoniemi, E., Himberg, J., Parhankangas, J., & Vesanto, J. (2002). SOM Toolbox - Online documentation, from http://www.cis.hut.fi/projects/somtoolbox/
4. Ampazis, N., Minis, I., Design of cellular manufacturing systems using Latent Semantic Indexing and Self Organizing Maps,Computational Management Science, Volume 1, Numbers 3-4 , Pages 275-292, 2004
5. Bu, W., Liu, Z., Tan, J.. Industrial robot layout based on operation sequence optimization, International Journal of Production Research, 47(15), 4125 – 4145, 2009
6. Burbidge J L, Production flow analysis, The Production Engineer, 42(12):742, 1963
7. Burbidge, J. L. Group Technology in Engineering Industry. London: Mechanical Engineering Publications, 1979
8. Chan, F. T. S., Lau, K. W., Chan, L. Y., Lo, V. H. Y.,. Cell formation problem with consideration of both intracellular and intercellular movements, International Journal of Production Research, 46(10), 2589 – 2620, 2008
9. Chang P.C., Liao T.W., Combining SOM and fuzzy rule base for flow time prediction in semiconductor manufacturing factory, Applied Soft Computing, Volume 6, Issue 2, Pages 198-206, 2006
10. Chattopadhyay M., Chattopadhyay S., Dan P. Machine-part cell formation through visual decipherable clustering of self-organizing map, The International Journal of Advanced Manufacturing Technology, Volume 52, Numbers 9-12, pp. 1019-1030, 2011
11. Choobineh, F.,. A framework for the design of cellular manufacturing systems. International Journal of Production Research, 26, 1161-1172, 1988
12. Chu, Chao-Hsien.,. Manufacturing cell formation by competitive learning, International Journal of Production Research, Volume 31, Issue 4 , pages 829 – 843, 1993
13. Corona F., Mulas M., Baratti R., Romagnoli J. A. On the topological modeling and analysis of industrial process data using the SOM, Computers & Chemical Engineering, Volume 34, Issue 12, 9, Pages 2022-2032, 2010
14. Dixit, R. A., Mishra, P.K. Cell formation considering real-life production parameters, International Journal of Manufacturing Technology and Management, 20(1-4), 197 – 221, 2010
15. Fernandez E.A., Balzarini M., Improving cluster visualization in self-organizing maps: Application in gene expression data analysis, Computers in Biology and Medicine, Volume 37, Issue 12, Pages 1677-1689, 2007
16. Flexer, A., On the use of self-organizing maps for clustering and visualization Intelligent Data Analysis, 5(5), 373 – 384, 2001
17. Fuertes J. J., Domínguez M., Reguera P., Prada M. A., Díaz I., Cuadrado A.A., Visual dynamic model based on self-organizing maps for supervision and fault detection in industrial processes, Engineering Applications of Artificial Intelligence, Volume 23, Issue 1, Pages 8-17, 2010
18. Ghaseminezhad M.H., Karami A., A novel self-organizing map (SOM) neural network for discrete groups of data clustering, Applied Soft Computing, Volume 11, Issue 4, Pages 3771-3778, 2011
19. Harhalakis, G., Nagi, R., Proth, J.M.,. An efficient heuristic in manufacturing cell formation for group technology applications, International Journal of Production Research,28(1), 185–198, 1990
20. Hautaniemi S.; Yli-Harja O.; Astola J.; Kauraniemi P.; Kallioniemi A.; Wolf M.; Ruiz J.7 Mousses S.; Kallioniemi O-P., Analysis and Visualization of Gene Expression Microarray Data in Human Cancer Using Self-Organizing Maps, Machine Learning, 52(1-2), , pp. 45-66(22), 2003
21. Haykin, S. Neural Networks, a Comprehensive Foundation, 2nd edition, Prentice Hall, 1999
22. Hébrail G., Hugueney B., Lechevallier Y., Rossi F., Exploratory analysis of functional data via clustering and optimal segmentation, Neurocomputing, Volume 73, Issues 7-9, Pages 1125-1141, 2010
23. Heikkinen M., Hiltunen T., Liukkonen M., Kettunen A., Kuivalainen R., and Hiltunen Y., A modelling and optimization system for fluidized bed power plants, Expert Systems with Applications, vol. 36, no. 7, pp. 10274–10279, 2009.

24. Inho, J., Jongtae, R.. Generalized machine cell formation considering material flow and plant layout using modified self-organizing feature maps, Computers & Industrial Engineering, 33(3-4), 457-460, 1997
25. Jämsä-Jounela S.-L., Vermasvuori M., Endén P., and Haavisto S., A process monitoring system based on the Kohonen self-organizing maps, Control Engineering Practice, vol. 11, no. 1, pp. 83–92, 2003.
26. Johnsson, M. and Balkenius, C., Haptic Perception with Self-Organizing ANNs and an Anthropomorphic Robot Hand, Journal of Robotics, Article ID 860790, 9 pages, doi:10.1155/2010/860790, 2010
27. Johnsson, M. and Balkenius, C., Sense of Touch in Robots With Self-Organizing Maps, Robotics, IEEE Transactions on, Volume: 27, Issue: 3, pages 498 – 507, 2011 doi: 10.1109/TRO.2011.2130090
28. Kaski S.: Data Exploration Using Self-Organizing Maps, Acta Polytechnica Scandinavia, 82, 1997.
29. Kaski, S., Kohonen, T., & Venna, J. Tips for SOM Processing and Colorcoding of Maps. In G. Deboeck & T. Kohonen (Eds.), *Visual explorations in finance with self-organizing maps* (pp. 195-202). New York: Springer-Verlag, 1998
30. Kasslin M., Kangas J., and Simula O., “Process State Monitoring Using self-organizing maps,” in Artificial Neural Networks 2, I. Aleksander and J. Taylor, Eds., vol. 1, North-Holland, Amsterdam, The Netherlands, 1992.
31. Khosravi M. H., Safabakhsh R., Human eye sclera detection and tracking using a modified time-adaptive self-organizing map, Pattern Recognition, Volume 41, Issue 8, Pages 2571-2593, 2008
32. Kiang, M.Y., Kulkarni, U.R. and Tam, K.Y., Self-organising map network as an interactive clustering tool – an application to group technology. Decision Support Systems, 15, pp.351-374, 1995
33. Kiang, Y. M., Extending the Kohonen self-organizing map networks for clustering analysis, Computational Statistics & Data Analysis, 38 (2001) 161–180, 2001
34. Kivilnoto, K.,. Topology preservation in self-organizing maps, In: Proceedings of ICNN’96, IEE International Conference on Neural Networks. IEEF, Service Center,Piscataway, pp. 294-299, 1996
35. Kohonen T.. Self-organization and associative memory. Springer, Heidelberg, 1984.
36. Kohonen, T.,. Proc IEEE 78,1464–1480, 1991
37. Kohonen T., Oja E., Simula O., Visa A., and Kangas J.. Engineering Applications of the Self-Organizing Map. Proceedings of the IEEE, 84(10):1358-1384, 1996.
38. Kohonen, T.,. Self-Organizing Maps (Springer, Berlin), 1997
39. Kohonen T., Proc. 8th Int. Conf.on Artificial Neural Networks (ICANN'98), L. Niklasson, M. Boden, and T. Ziemke (eds.), Springer, London, 1998, p. 65.
40. Kohonen T., *self-organizing maps*, Springer, Berlin, Germany, 2001
41. Kumar L. and Jain, P.K., Part-Machine Group Formation with Operation Sequence, Time and Production Volume, Int J Simul Model, 7(4), Pages 198-209, 2008
42. Kumar, C.S., Chandrasekharan, M.P.,. Grouping efficacy: a quantitative criterion for goodness of block diagonal forms of binary matrices in group technology, International Journal of Production Research, 28 (2), 233-243, 1990
43. Kuo R. J., Chi S. C., Teng P. W., Generalized part family formation through fuzzy self-organizing feature map neural network, Computers & Industrial Engineering, Volume 40, Issues 1-2, Pages 79-100, 2001
44. Lampinen J. and Oja E.. Clustering properties of hierarchical self-organizing maps, Journal of Mathematical Imaging and Vision, 2(2-3), Pages 261-272, 1992.
45. Liukkonen M., Havia E., Leinonen H., and Hiltunen Y., “Quality-oriented optimization of wave soldering process by using self-organizing maps,” Applied Soft Computing, vol. 11, no. 1, pp. 214–220, 2011.
46. Liukkonen M., Havia E., Leinonen H., and Hiltunen Y., Application of self-organizing maps in analysis of wave soldering process, Expert Systems with Applications, vol. 36, no. 3, pp. 4604–4609, 2009.
47. Mahdavi I., and Mahadevan B., CLASS: An algorithm for cellular manufacturing system and layout design using sequence data, Robotics and Computer-Integrated Manufacturing archive, Volume 24 , Issue 3, Pages: 488-497, 2008

48. Mahdavi I., Shirazi B. and Paydar M. M., A flow matrix-based heuristic algorithm for cell formation and layout design in cellular manufacturing system, THE INTERNATIONAL JOURNAL OF ADVANCED MANUFACTURING TECHNOLOGY, Volume 39, Numbers 9-10, 943-953, DOI: 10.1007/s00170-007-1274-7, 2008
49. Mari M., Nadal M., Schuhmacher M. and Domingo J. L., Application of Self-Organizing Maps for PCDD/F Pattern Recognition of Environmental and Biological Samples to Evaluate the Impact of a Hazardous Waste Incinerator, Environ. Sci. Technol., 44 (8), pp 3162–3168, 2010, DOI: 10.1021/es1000165
50. Mitrafanov, S.P.,. The Scientific Principles of Group Technology, National Lending Library Translation, UK, 1966
51. Moschou V., Ververidis D., Kotropoulos C., Assessment of self-organizing map variants for clustering with application to redistribution of emotional speech patterns, Neurocomputing, Volume 71, Issues 1-3, December, Pages 147-156, 2007
52. Oja M., Kaski S., and Kohonen T., Bibliography of self-organizing map (SOM) papers: 1998–2001 addendum, *Neural Computing Surveys*, vol. 3, pp. 1–156, 2002.
53. Pandian, S. R., Mahapatra, S.S., Cell formation with ordinal-level data using ART1-based neural networks, Int. J. Services and Operations Management, 4(5), 628-630, 2008
54. Pandian, S. R., Mahapatra, S.S., Manufacturing cell formation with production data using neural networks, Computers & Industrial Engineering, 56(4), 1340-1347, 2009
55. Pandit Y.P., Badhe Y. P., Sharma B.K., Tambe S. S., Kulkarni B. D., Classification of Indian power coals using K-means clustering and Self Organizing Map neural network, Fuel, Volume 90, Issue 1, , Pages 339-347, 2011
56. Papaioannou G., Wilson J. M., The evolution of cell formation problem methodologies based on recent studies (1997–2008): Review and directions for future research, European Journal of Operational Research, Volume 206, Issue 3, Pages 509-521, 2010
57. Pölzlbauer G. Survey and comparison of quality measures for selforganizing maps. In J´an Parali˘c, Georg P¨olzlbauer, and Andreas Rauber, editors,Proceedings of the Fifth Workshop on Data Analysis (WDA'04), pages 67–82, Sliezsky dom, Vysok´e Tatry, Slovakia, June 24–27 2004. Elfa Academic Press.
58. Ravi V., Kurniawan H., Thai P.N. K., Kumar, P. R., Soft computing system for bank performance prediction, Applied Soft Computing, Volume 8, Issue 1, Pages 305-315, 2008
59. Rezaeian J., Javadian N., Tavakkoli-Moghaddam R., Jolai F., A hybrid approach based on the genetic algorithm and neural network to design an incremental cellular manufacturing system, Applied Soft Computing, Volume 11, Issue 6, Pages 4195-4202, 2011
60. Sarker, B., Xu, Y. R., Operation sequences-based cell formation methods: A critical survey, Production Planning & Control, 9(8), 771 – 783, 1998
61. Singh S. P. and Sharma R. R. K. A review of different approaches to the facility layout problems, THE INTERNATIONAL JOURNAL OF ADVANCED MANUFACTURING TECHNOLOGY, Volume 30, Numbers 5-6, 425-433, DOI: 10.1007/s00170-005-0087-9, 2006
62. Skupin, A., & Agarwal, P. What is a Self-organizing Map? In P. Agarwal & A. Skupin (Eds.), Self-Organising Maps: applications in geographic information science (pp. 1-20). Chichester, England: John Wiley & Sons, 2008
63. Srinivasulu, S. , Jain, A. A comparative analysis of training methods for artificial neural network rainfall-runoff models, Applied Soft Computing, Volume 6, Issue 3, March 2006, Pages 295-306, 2006
64. Suresh N. C., Slomp J. and Kaparthi S., Sequence-Dependent Clustering of Parts and machines:a Fuzzy ART neural network approach, Iint. J. Prod. Res., 1999, vol. 37, no. 12, 2793- 2816, 1999
65. Tam, K. Y., An operation sequence based similarity coefficient for part families formations, Journal of Manufacturing Systems, 9(1), 55-68, 1990
66. Tas¸ demir, K., Exploring topology preservation of SOMs with a graph based visualization. Lecture Notes in Computer Science 5326, 180–187, 2008
67. Tompkins, J.A., White, J.A., Bozer, Y.A., Frazelle, E.H., Tanchoco, J.M.A., Trevino, J., 1996. Facilities Planning. Wiley, New York.
68. Ultsch, A. Maps for the Visualization of high-dimensional Data Spaces. In *Proceedings Workshop on Self-Organizing Maps* (pp. 225-230). Kyushu, Japan, 2003
69. Ultsch, A., & Siemon, H. Exploratory data analysis: Using Kohonen networks on transputers. Technical Report 329, University of Dortmund, Germany, 1989

70. Vakharia, A. J., Wemmerlov, U., Designing a Cellular Manufacturing System: A Materials Flow Approach Based on Operation Sequences, IIE Transactions, 22(1), 84 – 97,1990
71. Venkumara, P., Haq, A. N., Complete and fractional cell formation using Kohonen self-organizing map networks in a cellular manufacturing system, International Journal of Production Research, 44(20), 4257 – 4271, 2006
72. Venugopal, V., Narendran, T. T., Machine-cell formation through neural network models, International Journal of Production Research, 32(9), 2105 – 2116, 1994
73. Vesanto J. Som-based data visualization methods. Intelligent Data Analysis, 3(2):111–126, 1999.
74. Vesanto J., Sulkava M., and Hollmen J., On the Decomposition of the Self-Organizing Map Distortion Measure, Proceedings of the Workshop on Self-Organizing Maps,WSOM'03, Pages 11-16, 2003
75. Vesanto, J., Alhoniemi, E.,. Clustering of the Self-Organizing Map, IEEE Transactions on Neural Networks, 11(3), 586–600, 2000
76. Wang, J. A linear assignment clustering algorithm based on the least similar cluster representatives. IEEE Transactions on Systems Man and Cybernetics Part A – Systems and Humans, 29(1), 100–104, 1999
77. Wemmerlöv U. Economic justification of group technology software: Documentation and analysis of current practices, Journal of Operations Management, Volume 9, Issue 4, 1990, Pages 500-525
78. Wemmerlov U.; Johnson D. J. Cellular manufacturing at 46 user plants: implementation experiences and performance improvements, International Journal of Production Research, Volume 35, Number 1, 1 January 1997 , pp. 29-49(21)
79. WEMMERLOV, U., and HYER, N. L., , Research issues in cellular manufacturing. *International Journal of Production Research*, 25, 413-431, 1987
80. Wilppu, R., The Visualisation Capability of Self-Organizing Maps to Detect Deviation in Distribution Control.TUCS Technical Report No. 153. Turku Centre for Computer Science, Finland, 1997
81. Yamada S., Recognizing environments from action sequences using self-organizing maps, Applied Soft Computing, Volume 4, Issue 1, Pages 35-47, 2004
82. Yin, Y., & Yasuda, K.. Similarity coefficient methods applied to the cell formation problem: A taxonomy and review. International Journal of Production Economics, 101(2), 329–352, 2006
83. Yu J., A hybrid feature selection scheme and self-organizing map model for machine health assessment, Applied Soft Computing, Volume 11, Issue 5, Pages 4041-4054, 2011
84. Zhang K., Chai Y., Yang S. X., Self-organizing feature map for cluster analysis in multi-disease diagnosis, Expert Systems with Applications, Volume 37, Issue 9, Pages 6359-6367, 2010

## Appendices

Table 1 Machine-Part Incidence Matrix based on operation sequence example#1(artificially generated)

| | *m*1 | *m*2 | *m*3 | *m*4 | *m*5 | *m*6 |
|---|---|---|---|---|---|---|
| p1 | 0 | 1 | 0 | 3 | 0 | 2 |
| p2 | 1 | 0 | 3 | 0 | 2 | 0 |
| p3 | 0 | 1 | 0 | 3 | 0 | 2 |
| p4 | 1 | 0 | 3 | 0 | 2 | 0 |
| p5 | 0 | 1 | 0 | 3 | 0 | 2 |
| p6 | 1 | 2 | 0 | 0 | 0 | 3 |
| p7 | 2 | 1 | 0 | 0 | 3 | 0 |
| p8 | 2 | 1 | 0 | 0 | 3 | 0 |

Table 2 Block Diagonal Form after cell formation of example#1

| | *m*2 | *m*6 | *m*4 | *m*1 | *m*5 | *m*3 |
|---|---|---|---|---|---|---|
| p3 | 1 | 2 | 3 | 0 | 0 | 0 |
| p6 | 2 | 3 | 0 | 1 | 0 | 0 |
| p1 | 1 | 2 | 3 | 0 | 0 | 0 |
| p5 | 1 | 2 | 3 | 0 | 0 | 0 |
| p2 | 0 | 0 | 0 | 1 | 2 | 3 |
| p8 | 1 | 0 | 0 | 2 | 3 | 0 |
| p4 | 0 | 0 | 0 | 1 | 2 | 3 |
| p7 | 1 | 0 | 0 | 2 | 3 | 0 |

Table 3. Comparative computational results obtained by the proposed SOM approach.

| problem# | References | size of problem | SOM Map size | GTE (best result from literature) | GTE (using proposed approach) | % Increase in performance |
|---|---|---|---|---|---|---|
| example#1 | artificially generated | 8X6 | 7x7 | - | 81.25 | - |
| 1. | Nair & Narendran (1998) | 7X7 | 7x7 | 100 | 100 | 0 |
| 2. | Nair & Narendran (1998) | 20X8 | 9x9 | 85.71 | 85.71 | 0 |
| 3. | Nair & Narendran (1998) | 20X20 | 9x9 | 64.3 | 64.3 | 0 |
| 4. | Nair & Narendran (1999) | 12X10 | 10x10 | 84.61 | 84.61 | 0 |
| 5. | Sofianopoulou (1999) | 5X4 | 12x12 | 69.25 | 92.31 | 33.3 |
| 6. | Won and Lee (2001) | 5X5 | 15x15 | 84 | 84.61 | 0.73 |
| 7. | Pandian and Mahapatra (2008) | 7X5 | 18x18 | 58.54 | 87.82 | 50.02 |
| 8. | Pandian and Mahapatra (2008) | 8X6 | 19x19 | 83.93 | 87.5 | 4.25 |
| 9. | Park and Suresh (2003) | 19X12 | 21x21 | 78 | 78.46 | 0.59 |
| 10. | Sofianopoulou (1999) | 20X12 | 24x20 | 74.58 | 81.82 | 9.71 |
| 11. | Nagi et al. (1990) | 20X20 | 25x25 | 94 | 95.91 | 2.03 |
| 12. | Pandian and Mahapatra (2008) | 30X15 | 25x25 | 76.71 | 78.89 | 2.84 |
| 13. | Pandian and Mahapatra (2008) | 37X20 | 32x32 | 71.59 | 82 | 14.54 |
| 14. | Pandian and Mahapatra (2008) | 50X25 | 36x36 | 81.2 | 83.64 | 3 |

Table 4 SOM quality measures of the all the 14 problem data sets for the optimized SOM map size: quantizatisation error (QE), topographic error (TE), and computation time (in seconds). The example#1 is the artificially generated example and 1 to 14 are available problems.

| **problem#** | **problem size** | **SOM map size** | **CPU** | **QE** | **TE** | **ADM** |
|---|---|---|---|---|---|---|
| example#1 | 8X6 | 7x7 | 2 | 0 | 0 | 0.393 |
| 1 | 7X7 | 12x12 | 11 | 0 | 0 | 0.0729 |
| 2 | 20X8 | 18x18 | 25 | 0 | 0 | 0.5251 |
| 3 | 20X20 | 24x20 | 77 | 0 | 0.05 | 1.2253 |
| 4 | 12X10 | 15x15 | 19 | 0 | 0 | 0.6495 |
| 5 | 5X4 | 7x7 | 4 | 0 | 0 | 0.247 |
| 6 | 5X5 | 9x9 | 8 | 0 | 0 | 0.3066 |
| 7 | 7X5 | 9x9 | 6 | 0 | 0 | 0.4603 |
| 8 | 8X6 | 12x12 | 11 | 0 | 0.125 | 11.5873 |
| 9 | 19X12 | 19x19 | 34 | 0 | 0.053 | 1.1654 |
| 10 | 20X12 | 21x21 | 33 | 0.001 | 0 | 39.2721 |
| 11 | 20X20 | 25x25 | 148 | 0 | 0 | 0.3802 |
| 12 | 30X15 | 25x25 | 88 | 0 | 0 | 48.1599 |
| 13 | 37X20 | 32x32 | 297 | 0 | 0 | 1.5185 |
| 14 | 50X25 | 36x36 | 491 | 0 | 0.02 | 2.6814 |

Table 5 Paired t-test for GTE of best result from literature with GTE using proposed approach

| | N | Mean | StDev | SE Mean |
|---|---|---|---|---|
| GTE(Best Result from literature) | 14 | 79.03 | 11.1185 | 2.9716 |
| GTE(Using Proposed Approach) | 14 | 84.8271 | 8.5107 | 2.2746 |
| Difference | 14 | -5.79714 | 9.22811 | 2.46632 |

95% CI for mean difference: (-11.12530, -0.46899)
T-Test of mean difference = 0 (vs not = 0): T-Value = -2.35 P-Value = 0.035

Table 6. Levene's test of equality of variances for GTE of best result from literature with GTE using proposed approach

| | |
|---|---|
| F (Observed value) | 1.411 |
| F (Critical value) | 4.225 |
| DF1 | 1 |
| DF2 | 26 |
| p-value (one-tailed) | 0.246 |
| alpha | 0.05 |

Test interpretation:

H0: The variances are identical.

Ha: At least one of the variances is different from another.

As the computed p-value is greater than the significance level alpha=0.05, one cannot reject the null hypothesis H0.
The risk to reject the null hypothesis H0 while it is true is 24.57%.

Table 7. Results of experiment of SOM quality measures for different map sizes

| problem# | problem size | SOM *map size* | *CPU* | *QE* | *TE* | *ADM* |
|---|---|---|---|---|---|---|
| example#1 | 8X6 | 7x7 | 2 | 0 | 0 | 0.393 |
| | | 10x10 | 5 | 0 | 0.375 | 0.0082 |
| | | 12x12 | 9 | 0 | 0 | 0.0155 |
| 1 | 7X7 | 7x7 | 3 | 0.024 | 0.571 | 2.5941 |
| | | 9x9 | 6 | 0.001 | 0.714 | 0.893 |
| | | 12x12 | 11 | 0 | 0 | 0.0729 |
| 2 | 20X8 | 15x15 | 11 | 0.001 | 0.05 | 1.7156 |
| | | 18x18 | 25 | 0 | 0 | 0.5251 |
| | | 20x20 | 41 | 0 | 0 | 0.2086 |
| 3 | 20X20 | 20x20 | 47 | 0 | 0.1 | 2.5226 |
| | | 24x20 | 77 | 0 | 0.05 | 1.2253 |
| | | 25x25 | 149 | 0 | 0.05 | 0.4314 |
| 4 | 12X10 | 10x12 | 6 | 0.012 | 0 | 3.9102 |
| | | 13x13 | 10 | 0.001 | 0.083 | 1.3721 |
| | | 15x15 | 19 | 0 | 0 | 0.6495 |
| 5 | 5X4 | 5x5 | 1 | 0.008 | 0 | 2.1125 |
| | | 7x7 | 4 | 0 | 0 | 0.247 |
| | | 12x12 | 17 | 0 | 0 | 0.0012 |
| 6 | 5X5 | 5x5 | 2 | 0.119 | 0 | 4.5958 |
| | | 7x7 | 4 | 0.005 | 0 | 1.6172 |
| | | 9x9 | 8 | 0 | 0 | 0.3066 |
| 7 | 7X5 | 7x7 | 3 | 0.004 | 0 | 1.46 |
| | | 9x9 | 6 | 0 | 0 | 0.4603 |
| | | 11x11 | 10 | 0 | 0 | 0.0485 |
| 8 | 8X6 | 8x8 | 3 | 0.044 | 0 | 11.5873 |
| | | 10x10 | 5 | 0.001 | 0 | 11.5873 |
| | | 12x12 | 11 | 0 | 0.125 | 11.5873 |
| 9 | 19X12 | 17x17 | 19 | 0.001 | 0 | 2.0344 |
| | | 19x19 | 34 | 0 | 0.053 | 1.1654 |
| | | 21x21 | 53 | 0 | 0.105 | 0.3056 |
| 10 | 20X12 | 15x15 | 11 | 0.034 | 0 | 39.2721 |
| | | 17x17 | 19 | 0.006 | 0 | 39.2721 |
| | | 21x21 | 33 | 0.001 | 0 | 39.2721 |
| 11 | 20X20 | 20x20 | 47 | 0.001 | 0 | 60.4631 |
| | | 23x23 | 95 | 0 | 0.05 | 60.4631 |
| | | 25x25 | 148 | 0 | 0 | 0.3802 |
| 12 | 30X15 | 20x20 | 32 | 0.026 | 0.033 | 8.7488 |
| | | 25x25 | 88 | 0 | 0 | 48.1599 |
| | | 30x30 | 240 | 0 | 0 | 0.5747 |
| 13 | 37X20 | 25x25 | 89 | 0.004 | 0 | 6.9443 |
| | | 29x29 | 176 | 0 | 0.027 | 3.2445 |
| | | 32x32 | 297 | 0 | 0 | 1.5185 |
| 14 | 50X25 | 30x30 | 165 | 0.005 | 0 | 7.6854 |
| | | 34x34 | 352 | 0.001 | 0.02 | 3.8743 |
| | | 36x36 | 491 | 0 | 0.02 | 2.6814 |

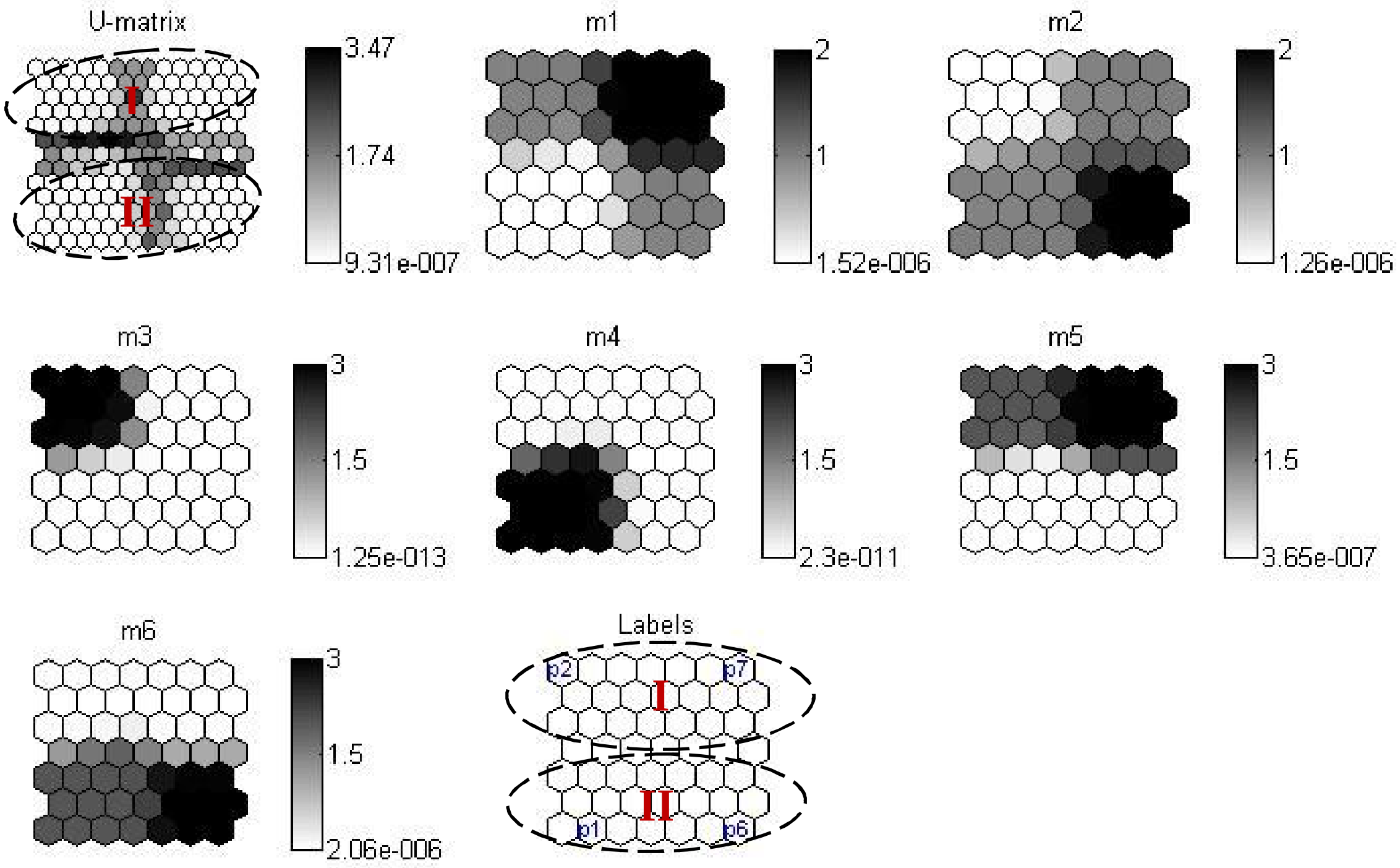

Figure 1. An example of representation SOM machine-part cluster visualization of the example#1 data. On topmost left is the U-matrix and on bottom right is the SOM map neuron label. The machine variable m1 through m6 are the component plane. The eight figures are associated by position. The SOM map size of the component is 7X7. The hexagon in a particular position in each figure corresponds to the identical map unit. The clusters can be interpreted straightforward from the distances between the nodes of U-matrix. Two circles in the U-matrix represent two clusters. Directly, machine variables can be compared using the component planes. For example, the map unit in upper half has low values for m2, m4, m6 and relatively high value for m1, m3, m5. The label associated with the map unit is p1, p3, p5, p6 and from the U-matrix it can be observed that the node is very close to its neighbors. Thus the cluster formed on the top half and the bottom half are designated as Cluster-I and Cluster-II respectively.

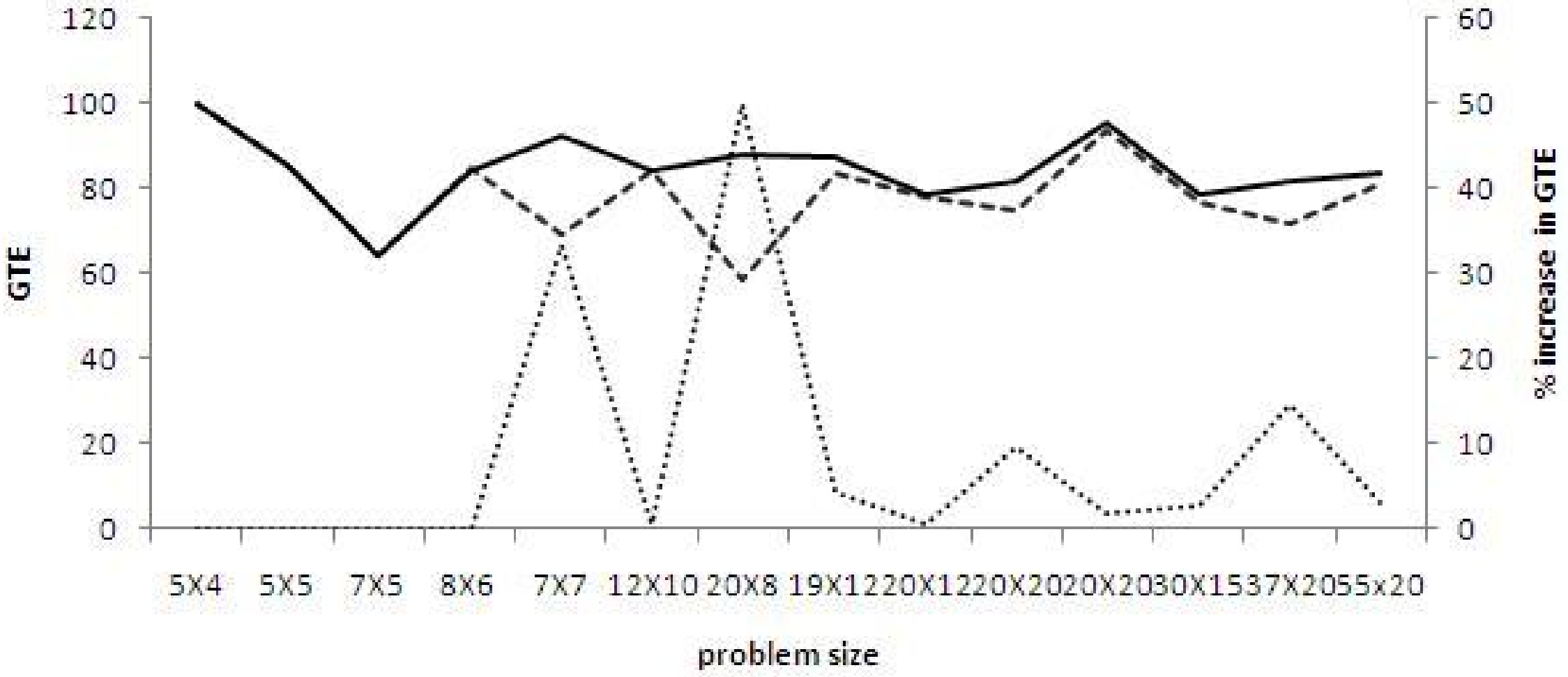

Figure 2. For measuring goodness of solution the output of GTE is shown higher in proposed solution when compared the same from literature based best solution in all the 14 problems (x-axis represents dimension of the 14 problems (solid line, thick dashed line and dotted line represent GTE obtained using proposed approach best GTE obtained from literature (left axis of GTE), and % increase in GTE(right axis as % increase in GTE).

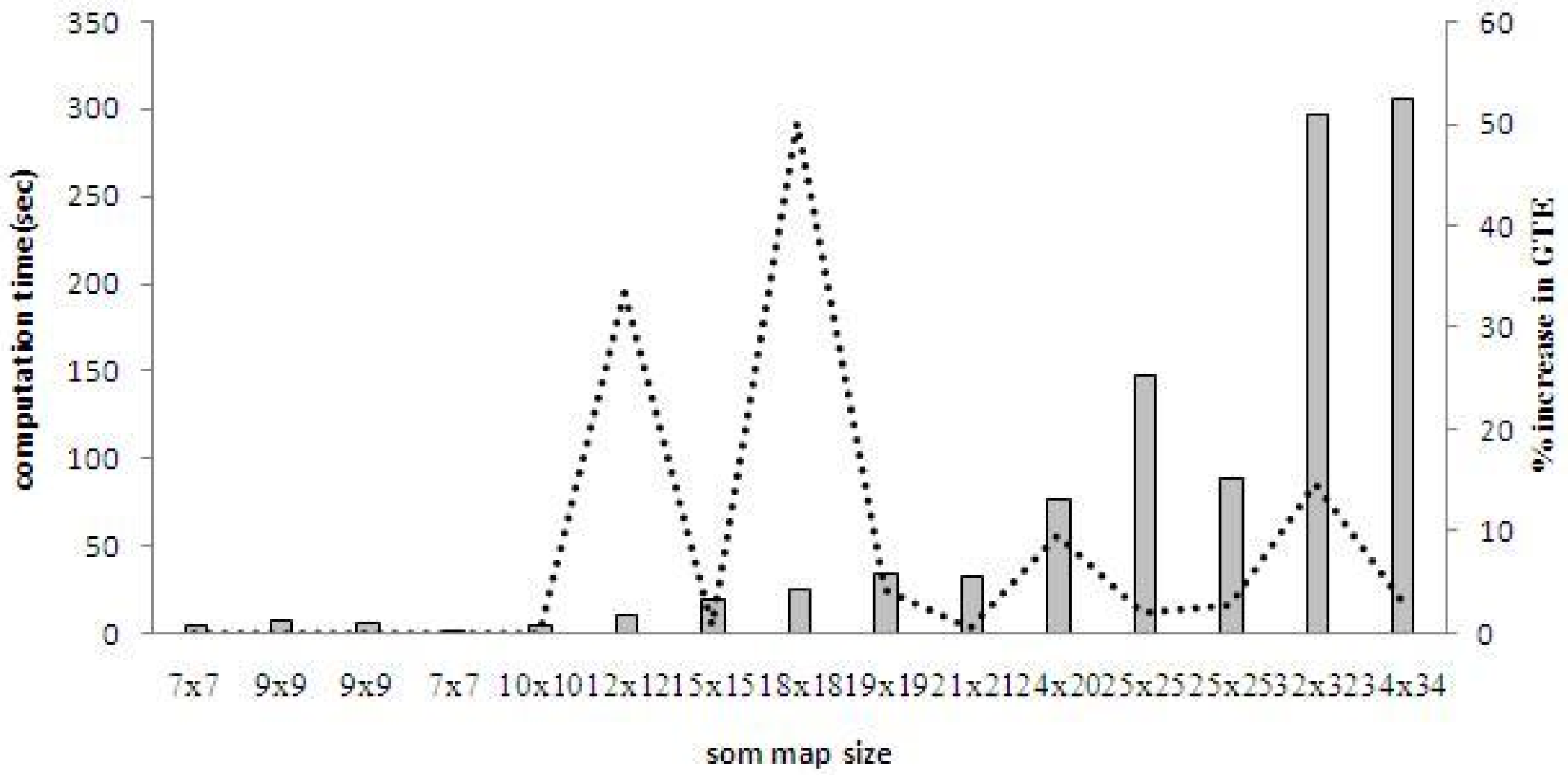

Figure 3. Comparison of % increase in GTE using proposed solution (represents dotted line in the right axis) with the computation time (represents grey scaled column in the left axis) taken to execute the proposed solution for all the different sizes of the 14 problem data sets (x-axis represents dimension of the SOM map). Even after higher increase in computation time there is still increase in performance (GTE) for the big size problem data sets. Average size problems (with average SOM map size) resulted with higher percentage increase in GTE in comparison with low computation time.

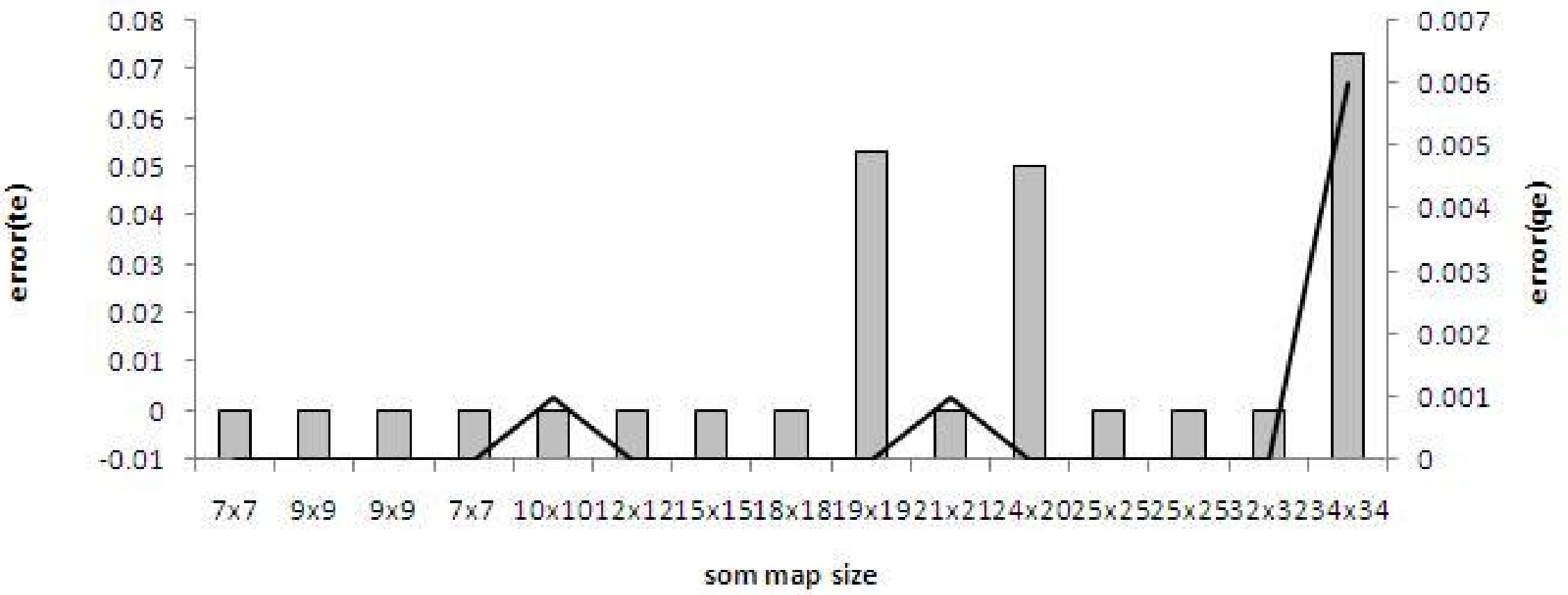

Figure 4. Grey colored vertical bar represents topography error (te) scaled in the left axis and solid line represents quantization error(qe) which is scaled in right axis. The x-axis represents som map size of all the problems. The both errors are highest for the problem# 15 with 55x20 dimension(som map size 34X34)

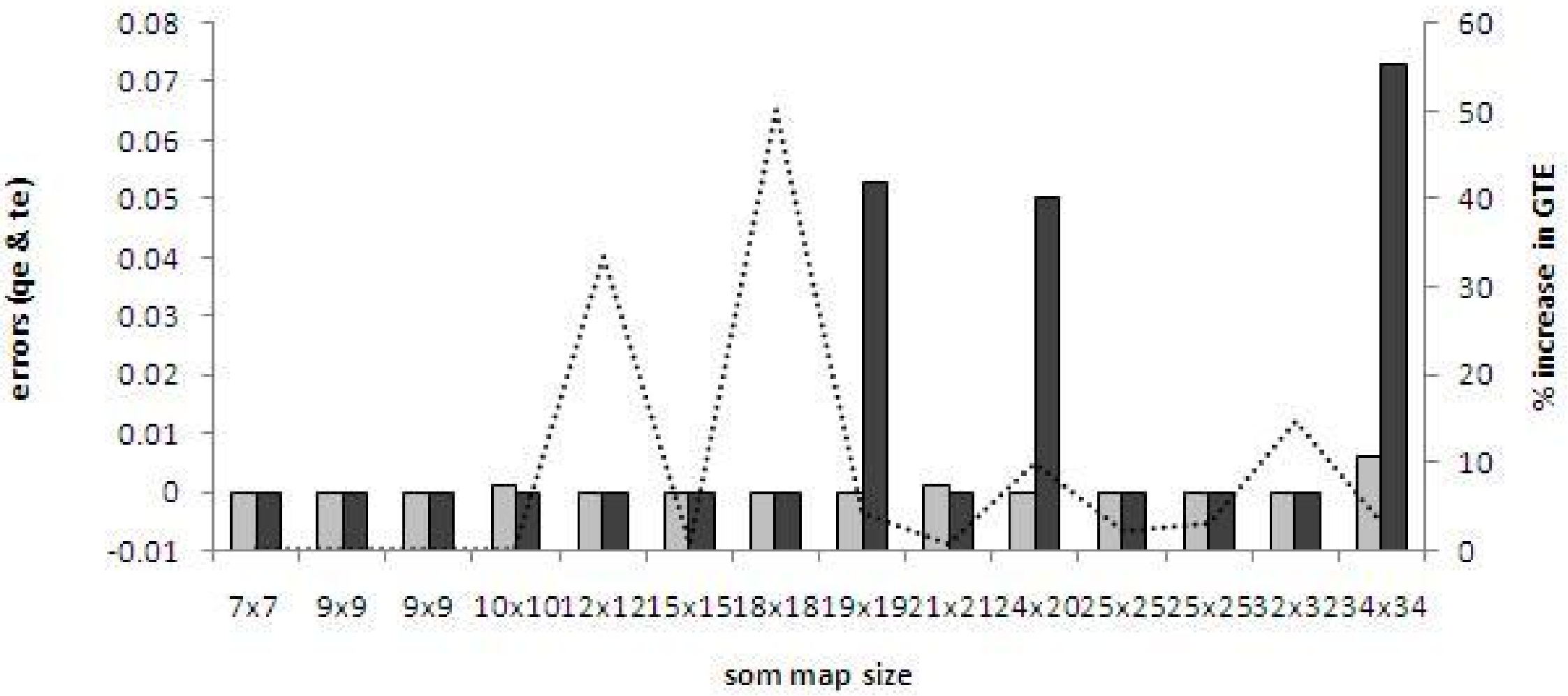

Figure 5. For majority of the optimum SOM map dimension the resulting QE and TE are zero with the corresponding % increase in GTE for all the 15 problem data sets (scaled in x-axis). Where black colored vertical bar and grey-scaled bar represent TE and QE (scaled in left axis). Dotted line represents % increase in GTE scaled in right axis.

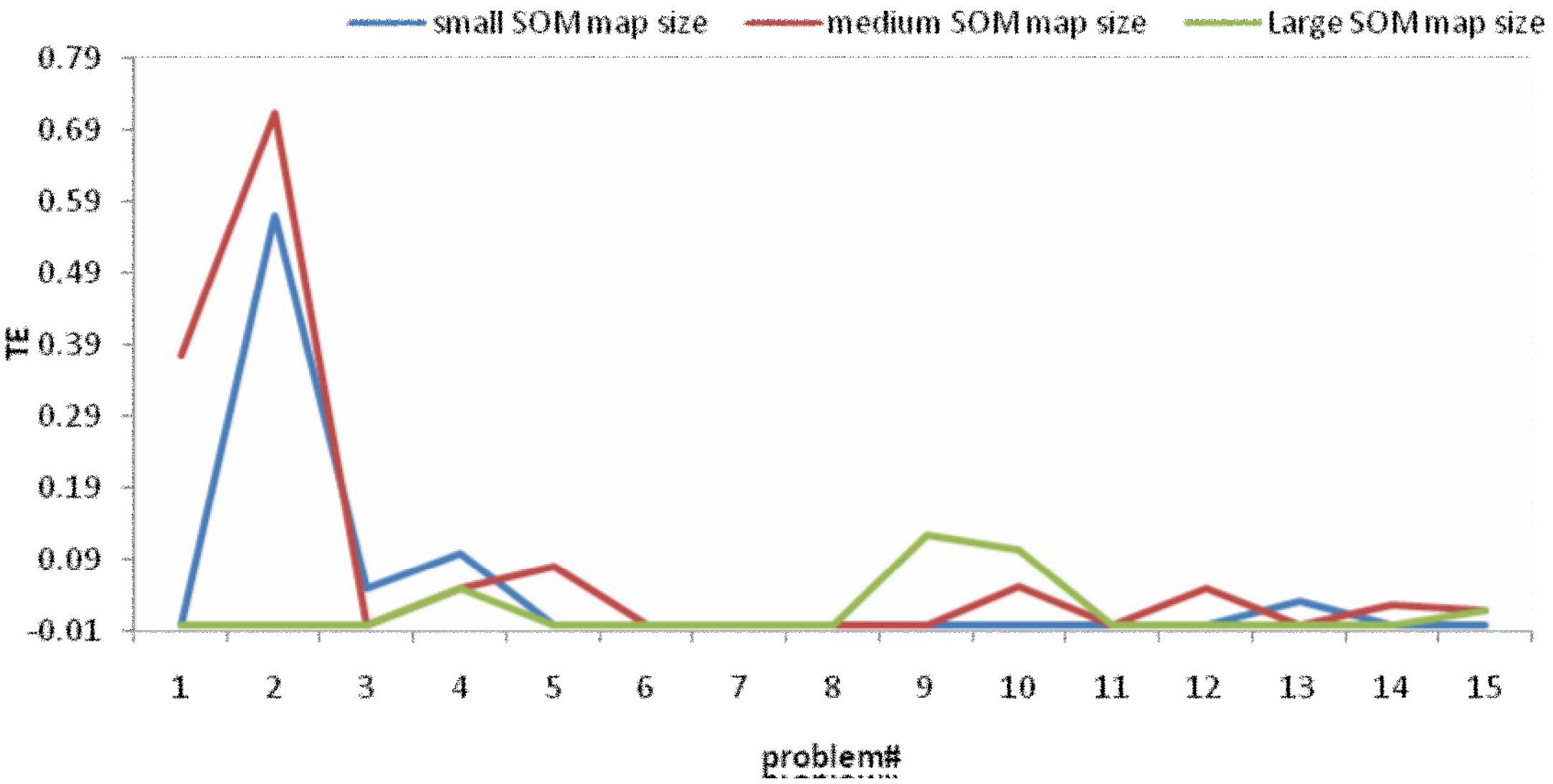

Figure 6. The pattern of TE value (scaled in y-axis extracted by all the 15 problem data sets (scaled in x-axis). The TE value gradually diminished for large SOM map size

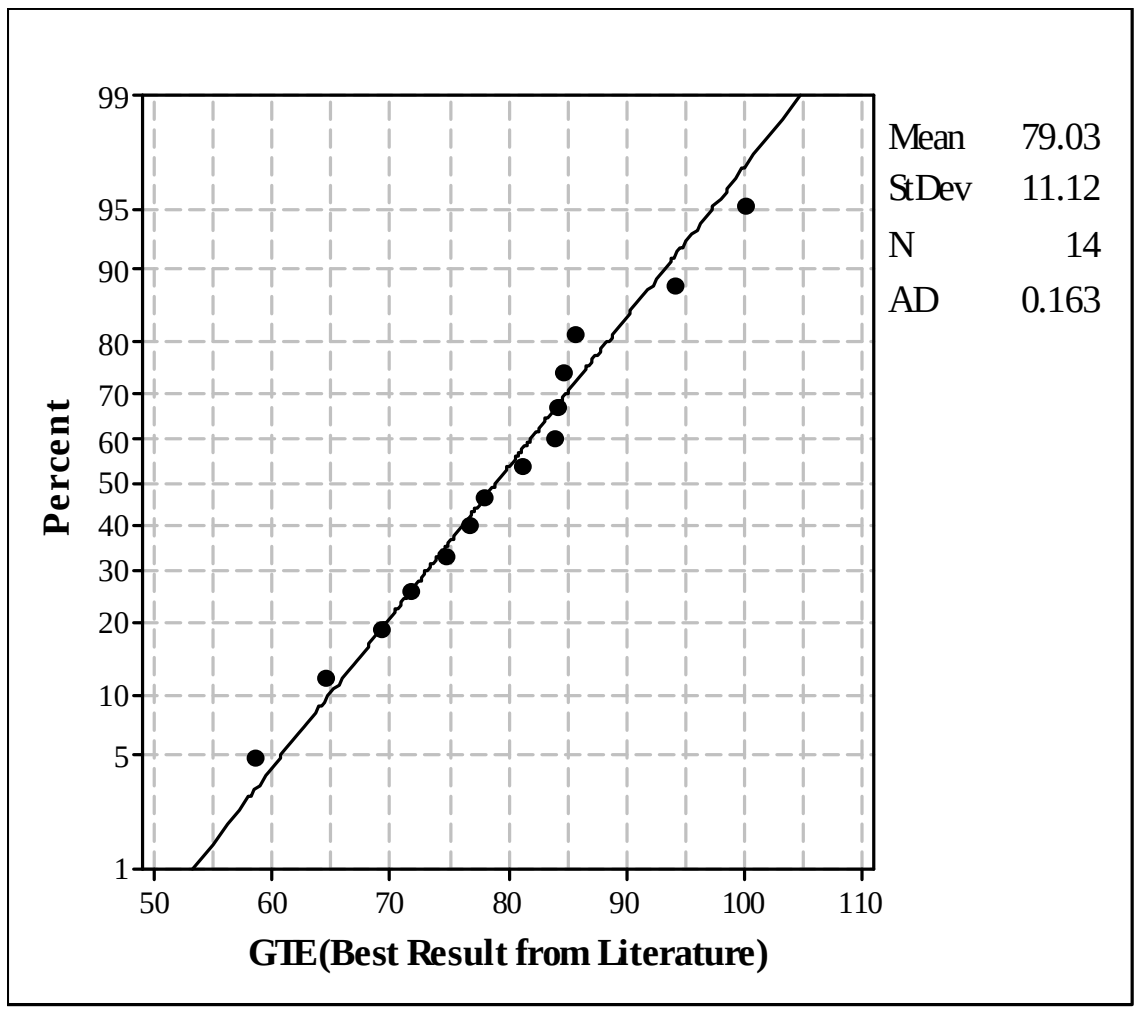

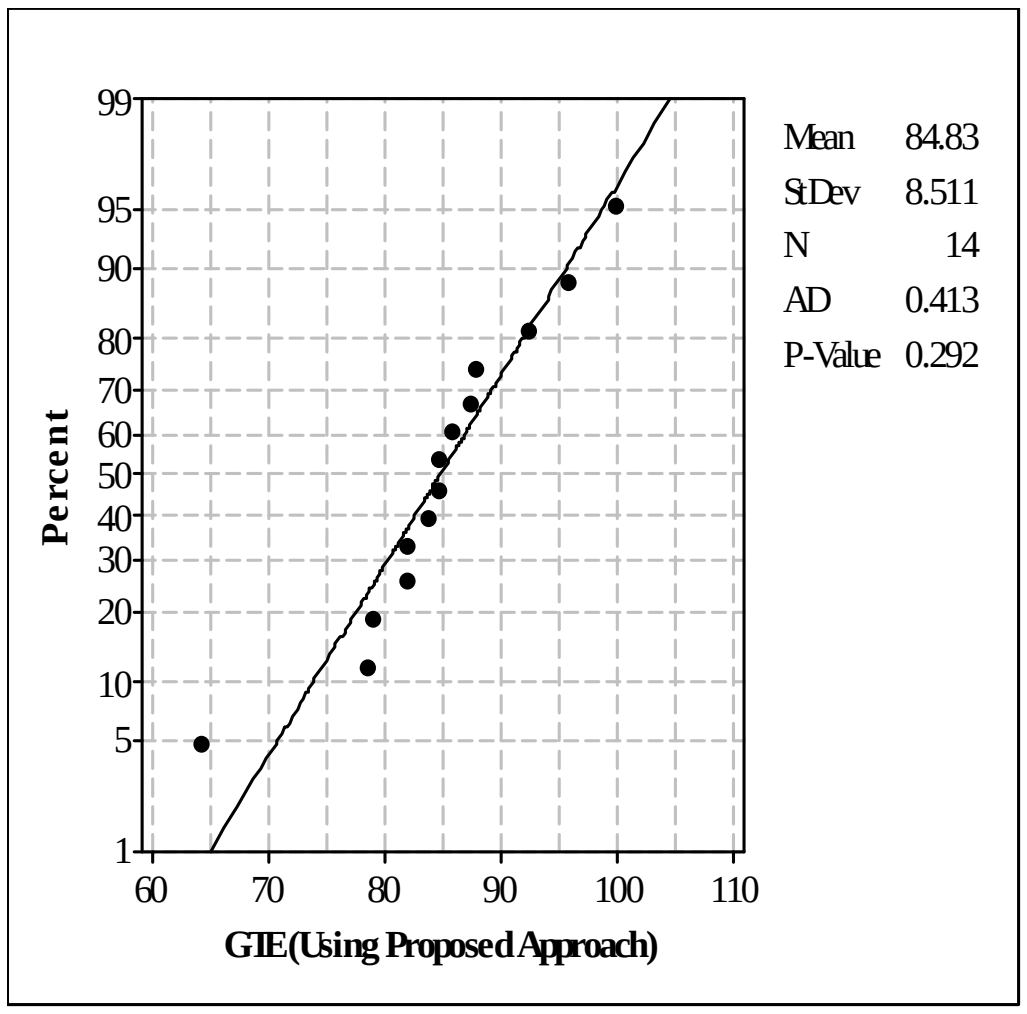

Figure 7 a. Probability Plot of GTE (Best Result from Literature)

Figure 7 b. Probability Plot of GTE (Using Proposed Approach)